%% file: lafurca-arxiv.tex
\documentclass{article}

\usepackage{algorithm, algpseudocode}
\usepackage{amsfonts, amsmath, amsopn, amsthm, epsfig}
  \numberwithin{equation}{section}
\usepackage{graphicx, epstopdf}
\usepackage{hyperref}
\usepackage{subfigure}
\usepackage{appendix}
\usepackage{authblk}

\usepackage{calc}
\usepackage[top=2.5cm,bottom=2.3cm,inner=2.3cm,outer=2.3cm,bindingoffset=0.5cm,footskip=0.55cm]{geometry}
\theoremstyle{remark}

\newenvironment{lemma*}[2][Lemma]{\par\bgroup{\bfseries #1\ #2. }\it\ignorespaces}{\egroup}

\title{LaFurca: Iterative Refined Speech Separation Based on Context-Aware Dual-Path Parallel Bi-LSTM}

\author[1]{Ziqiang Shi}
\author[1]{Rujie Liu}
\author[2]{Jiqing Han}

\affil[1]{Fujitsu Research and Development Center, Beijing, China}
\affil[2]{Harbin Institute of Technology, Harbin, China}

\input{macros.tex}

\date{}

\begin{document}

\maketitle

\renewcommand{\thefootnote}{\fnsymbol{footnote}}


\begin{abstract}
  Deep neural network with dual-path bi-directional long short-term memory (BiLSTM) block has been proved to be very effective in sequence modeling, 
  especially in speech separation, e.g. DPRNN-TasNet~\cite{luo2019dual}. In this paper, we propose several 
  improvements of dual-path BiLSTM based network for end-to-end approach to  monaural speech separation. Firstly a 
  dual-path network with intra-parallel BiLSTM and inter-parallel BiLSTM components is introduced to reduce performance sub-variances 
  among different branches. Secondly, we propose to use global context aware inter-intra cross-parallel BiLSTM to further perceive the global contextual 
  information. Finally, a spiral multi-stage  dual-path BiLSTM is proposed to iteratively refine the separation results of the previous stages. 
  All these networks take the mixed utterance of two speakers and map it to two separate utterances, where each utterance contains only one speaker's voice.  
  For the objective, we propose to train the network by directly optimizing the utterance level scale-invariant signal-to-distortion ratio (SI-SDR) in a 
  ermutation invariant training (PIT) style. Our experiments on the public WSJ0-2mix data corpus results in 20.55dB SDR improvement, 20.35dB SI-SDR improvement, 
  3.69 of PESQ, and 94.86\% of ESTOI, which shows our proposed networks can lead to performance improvement on the speaker separation task. 
  We have open-sourced our re-implementation of the DPRNN-TasNet in https://github.com/ShiZiqiang/dual-path-RNNs-DPRNNs-based-speech-separation, 
  and our LaFurca is realized based on this implementation of DPRNN-TasNet, it is believed that the results in this paper can be reproduced with ease.
  \end{abstract}

  \section{Introduction}
  \label{sec:introduction}

  Multi-talker monaural speech separation has a vast range of applications.
  For example, a home environment or a conference environment in which many people talk, the human auditory system can 
  easily track and follow a target speaker's voice from the multi-talker's mixed voice.
  In this case, a clean speech signal of the target speaker needs to be separated from the mixed speech to complete the subsequent 
  recognition work.
  Thus it is a problem that must be solved in order to achieve satisfactory performance in speech or speaker recognition tasks. 
  There are two difficulties in this problem, the first is that since we don't have any prior information of the user, a 
  practical system must be speaker-independent. The second difficulty is that there is no way to use the beamforming algorithm 
  for a single microphone signal. Many traditional methods, such as computational auditory scene analysis
   (CASA)~\cite{wang2006computational,shao2006model,hu2013unsupervised}, Non-negative matrix factorization 
  (NMF)~\cite{smaragdis2007convolutive, le2015sparse}, and probabilistic models~\cite{virtanen2006speech}, 
  do not solve these two difficulties well.
  
  Recently, a large number of techniques based on deep learning are proposed for this task. These methods can be briefly grouped into two categories: time-frequency (TF) domain methods (non-end-to-end) and time-domain methods (end-to-end). The first category is to use short-time Fourier transform (STFT) to decompose the time-domain mixture into the time-frequency domain to display and to separate therein. Usually, deep neural networks (DNN) is introduced for estimating the ideal binary or ratio masks (IBM or IRM), or phase-sensitive masks (PSM), and the source separation is transformed into a magnitude domain TF unit-level classification or regression problem, and mixed phases are usually retained for resynthesis. Notable work includes deep clustering (DPCL)~\cite{hershey2016deep,isik2016single}, permutation invariant training (PIT)~\cite{yu2017permutation}, and combinations of DPCL and PIT, such as Deep CASA~\cite{liu2019divide} and Wang et al.~\cite{wang2019deep}.The second category is end-to-end speech separation in time-domain~\cite{luo2017tasnet,luo2018tasnet,venkataramani2017adaptive,shi2019end, shi2019deep, zhang2020furcanext,luo2019dual,zeghidour2020wavesplit,nachmani2020voice}, which is a natural way to overcome the obstacles of the upper bound source-to-distortion ratio improvement (SDRi) in STFT mask estimation based methods and real-time processing requirements in actual use.
  
  This paper is based on the end-to-end 
  method~\cite{luo2017tasnet,luo2018tasnet,venkataramani2017adaptive,shi2019end, shi2019deep, zhang2020furcanext,luo2019dual,zeghidour2020wavesplit,nachmani2020voice}, which has achieved better 
  results than DPCL based or PIT based approaches. Since most DPCL and PIT based methods use STFT as front-end. Specifically, the mixed 
  speech signal is first transformed from one-dimensional signal in the time domain to two-dimensional spectrum signal in TF domain, and then 
  the mixed spectrum is separated to result in spectrums corresponding to different source speeches by a deep clustering or mask 
  estimation method, and finally, the cleaned source speech signal can be restored by an inverse STFT on each spectrum. This framework 
  has several limitations. Firstly, it is unclear whether the STFT  is optimal (even assume the parameters it depends on are optimal, 
  such as size and overlap of audio frames, window type, and so on) transformation of the signal for speech separation~\cite{shi2019cqt}. 
  Secondly, most STFT 
  based methods often assumed that the phase of the separated signal to be equal to the mixture phase, which is generally incorrect and 
  imposes an obvious upper bound on separation performance by using the ideal masks. As an approach to overcome the above problems, several
  speech separation models were recently proposed that operate
  directly on time-domain speech signals~\cite{luo2017tasnet,luo2018tasnet,venkataramani2017adaptive,shi2019end, shi2019deep, zhang2020furcanext,luo2019dual,zeghidour2020wavesplit,nachmani2020voice}.
  Inspired by these first results, we propose LaFurca\footnotemark[1], which is a general name for a series of fully end-to-end time-domain separation 
  methods, includes 1) dual-path network with intra- and inter-parallel bi-directional long short-term memory (BiLSTM)
  components: replace
  intra- and inter-BiLSTM~\cite{luo2019dual} by multiple parallel BiLSTM modules, which can reduce the variance of this model. The intra- and inter-paralleled BiLSTM modules replicate weight matrices and take the 
  average from the feature maps produced by those layers. This convenient technique can effectively improve separation performance; 
  2) global context aware inter-intra cross-parallel BiLSTM: In order to further perceiving the global contextual 
  information, intra- and intra-BiLSTM are placed side by side for mutual reference; 3) multiple spiral iterative refinement dual-path BiLSTM: inspired by~\cite{isik2016single,kavalerov2019universal}, in which the signal estimates from an initial 
  mask-based separation network serves as input, along with the original mixture, to a second identical separation network.

  \footnotetext[1]{`LaFurca' is `the fork', where La is from French, and Furca is from Latin. It is used to mean that mixed speech is divided into two streams by our network, just like water.}
  
  The remainder of this paper is organized as follows:  section~\ref{sec:dplstm} introduces end-to-end monaural
  speech separation based on deep neural networks with dual-path  BiLSTM blocks. 
  Section~\ref{sec:lafurca} describe our proposed 
  LaFurca and the 
  separation algorithm in detail. 
  The experimental setup and results are presented in Section~\ref{sec:experiments}. We conclude this paper in Section~\ref{sec:conclusion}.
  
  \section{Speech separation with dual-path BiLSTM blocks}
  \label{sec:dplstm}
  
  \begin{figure}
    \centering
    \hspace{-5mm}
    \includegraphics[width=1.0\linewidth]{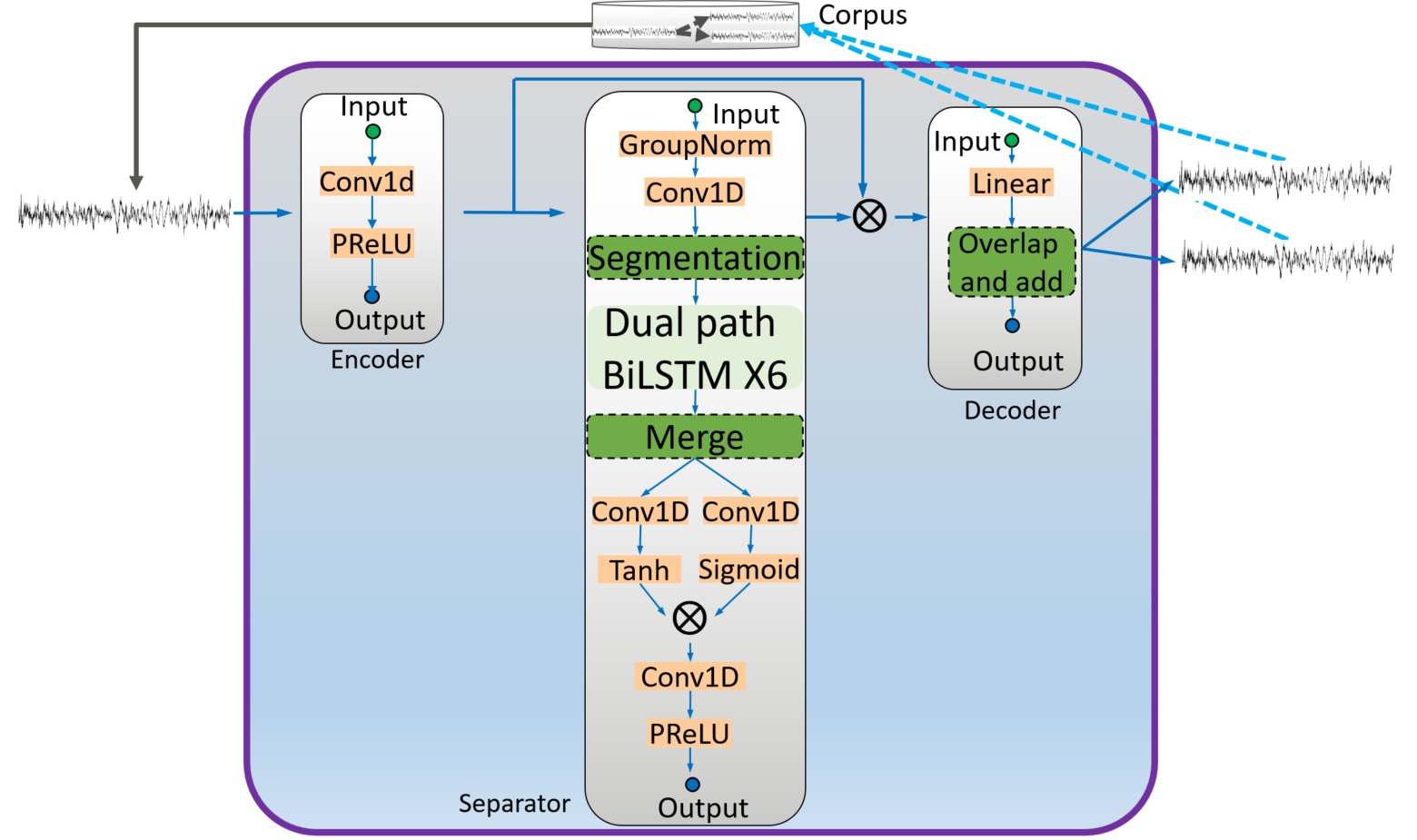}
    \hspace{-5mm}
    \caption{
    The pipeline of dual-path BiLSTM based speech separation in~\cite{luo2019dual}, 
    which is called DPRNN-TasNet.
    }
    \label{dprnn_pipeline}
    \end{figure}
  
  In this section, we review the formal definition of the monaural speech separation task and the original 
  dual-path BiLSTM based 
  separation architecture~\cite{luo2019dual}.
  
  The goal of monaural speech separation is to estimate the individual target signals from a linearly mixed single-microphone signal, 
  in which the target signals overlap in the TF domain.
  Let $x_i(t),i=1,..,S$ denote the $S$ target speech signals and  $y(t)$ denotes the
  mixed speech respectively. If we assume the target signals are linearly mixed, which can be represented as:
  \begin{equation*}
  y(t)=\sum_{i=1}^{S}x_i(t),
  \end{equation*}
  then monaural speech separation aims at estimating individual target signals from
  given mixed speech $y(t)$. In this work it is assumed that the number of target signals is known.
  
  In order to deal with this ill-posed problem, Luo et al.~\cite{luo2018tasnet,luo2019dual} 
  introduce adaptive front-end methods to achieve high speech separation performance on WSJ0-2mix 
  dataset~\cite{hershey2016deep,isik2016single}. Such methods contain three processing stages, where 
  the state-of-the-art architecture~\cite{luo2019dual} is used as an illustration.
  As shown in 
  Figure~\ref{dprnn_pipeline}, the architecture consists of an encoder 1-D convolution (Conv1D in the Figure~\ref{dprnn_pipeline} for abbreviation 
  and the following description is in the same way and will not be repeated again) is followed by a parametric ReLU (PReLU), 
  a separator 
  (consisted in the order by a 
  layer normalization~\cite{ba2016layer} (LayerNorm), a 1$\times$1 convolution (1$\times$1conv), 
  6 dual-path BiLSTM layers, 1$\times$1 convolution, and a softmax operation) and a decoder of a fully connected (FC) layer.  First, 
  the encoder module is used to convert short 
  segments of the mixed waveform into their corresponding representations. Then, the representation is used 
  to estimate the multiplication 
  function (mask) of each source and each encoder output for each time step. The source waveform is then 
  reconstructed by transforming the
   masked encoder features using a linear decoder module. This framework is called DPRNN-TasNet 
   in~\cite{luo2019dual}.

  \begin{figure}[th]
  \centering
  \subfigure[The operation of `Segmentation'.]{
  \begin{minipage}[t]{1.0\linewidth}
  \centering
  \includegraphics[width=2.3in]{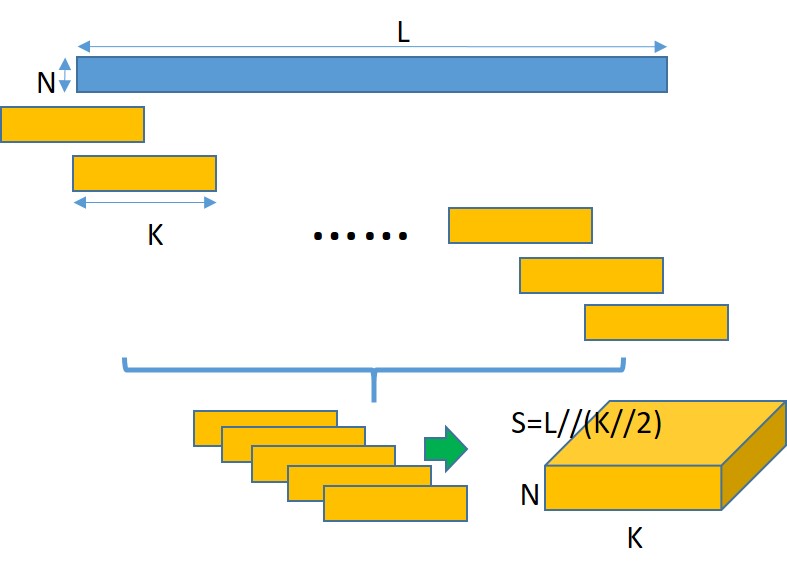}
  \label{segmentation}
  \end{minipage}%
  }%
  
  \subfigure[The structure of dual-path BiLSTM.]{
  \begin{minipage}[t]{1.0\linewidth}
  \centering
  \includegraphics[width=3.0in]{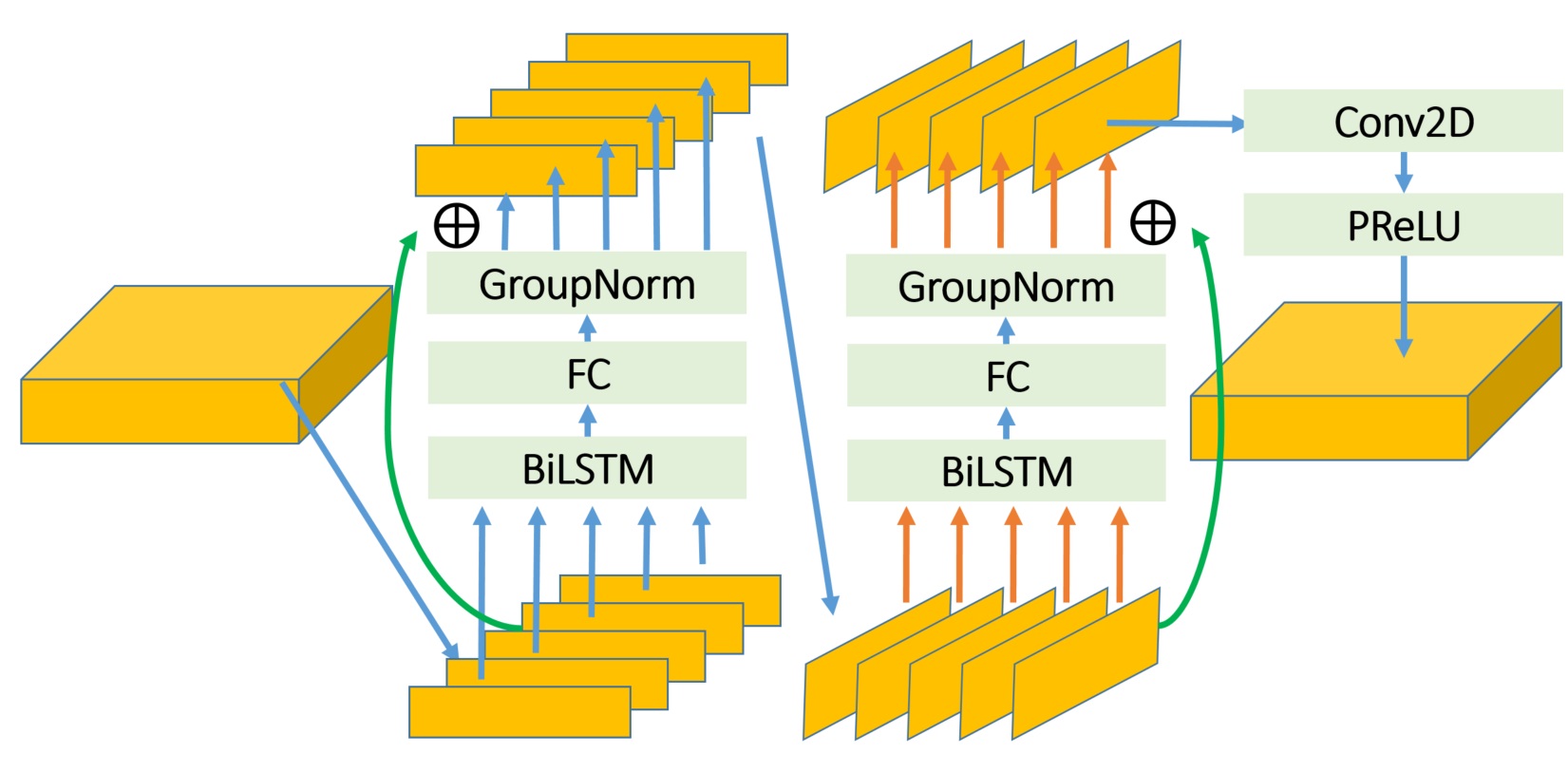}
  \label{dprnn}
  \end{minipage}%
  }%
  
  \subfigure[The operation of `Merge'.]{
  \begin{minipage}[t]{1.0\linewidth}
  \centering
  \includegraphics[width=2.3in]{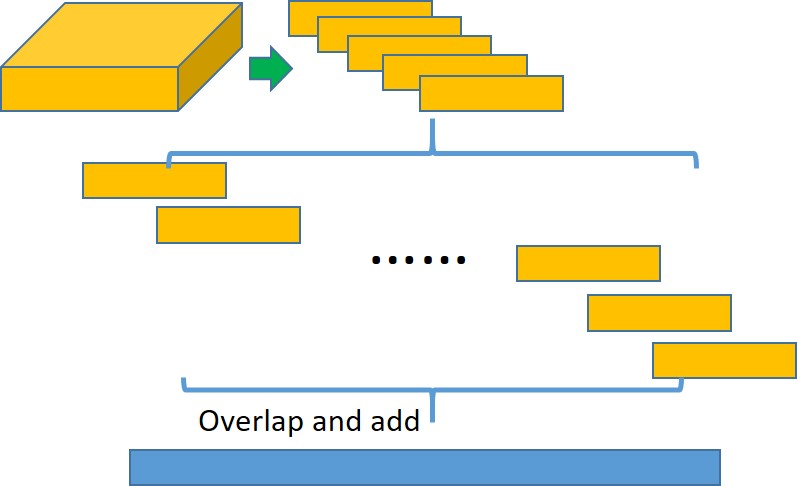}
  \label{Merge}
  \end{minipage}
  }%
  \centering
  \caption{Key components in the pipeline of DPRNN-TasNet}
  \end{figure}

  The key factors for the best performance of DPRNN-TasNet on WSJ0-2mix 
  dataset~\cite{hershey2016deep,isik2016single} are the local and global data chunk formulation
  and the dual-path BiLSTM module~\cite{luo2019dual}.
  Luo et al.~\cite{luo2019dual} first splits the output of the encoder into chunks with or without overlaps
  and concatenates them to form a 3-D tensor, as shown in Figure~\ref{segmentation}. 
  The dual-path BiLSTM 
  modules will map these 3-D tensors to 3-D tensor masks, as shown in Figure~\ref{dprnn}. The output 
  3-D tensor masks and the original 3-D tensor are converted back to a sequential output by a `Merge'
  operation as shown in Figure~\ref{Merge}.
  From Figure~\ref{segmentation} we can see the transformation from a long stream wav TF representation (although there is no frequency in the representation, since the representation is the output of the encoder, but here we use frequency for clarity) into a 3-D tensor. Here in Figure~\ref{segmentation},  the long stream TF representation is splitted into several segments of the same length, then these segments are concatenated together into a 3-D tensor. Thus the 3-D tensor has three axes, which are frequency, small-scale time, and large-scale time. For example, if we treat the 3-D tensor as several matrixes, the first matrix is from time 1 to time 100, the second matrix is from time 101 to time 200, etc. Then from other dimensions, the time indexes of the first column of the first matrix and the first column of the second matrix are 1 and 101, etc. For both the LSTMs, they process along the time axis, is not along the frequency axis.  The 3-D tensor as an instantaneous input to the LSTMs, which means all the small segments are separated pass through the intra-BiLSTM, for example, for the first matrix is from time 1 to time 100. The inter-BiLSTM works along the other dimension, for example, time 1, 101, 201, etc.

  Some architectures similar to dual-path BiLSTM have been proposed as alternatives to 
  the recurrent neural network (RNN) in various 
  tasks~\cite{zhang2018spatial,liu2017global}.  
  dual-path BiLSTM can organize any type of RNN layer and model long sequence inputs in a very simple way. 
  The intuition is to divide the input sequence into shorter blocks and interleave two BiLSTMs, intra-BiLSTM 
  and an inter-BiLSTM, for local and global modeling, respectively. In a dual-path BiLSTM, the intra-BiLSTM 
  first 
  processes the local block independently, and then the inter-BiLSTM summarizes the information from all the 
  blocks to 
  perform sound level processing.

  As shown in Figure~\ref{dprnn}, the input of intra-BiLSTM is a segment composed of several consecutive 
  frames in time, and an utterance is divided into several such segments. These segments are passed through 
  a BiLSTM, a fully connected projection, and a group normalization (GroupNorm in Figure~\ref{dprnn})~\cite{wu2018group} 
  operation respectively 
  A residue connection is added to the output of the group normalization to result in the final output of 
  the intra-BiLSTM with the same 
  shape as the input. The output of intra-BiLSTM will be used as the input of inter-BiLSTM, 
  but a permutation will be performed on this input to let  inter-BiLSTM capture global dependency.
  Here `permutation' means for the intra-BiLSTM, the input time index is the natural one, e.g. time 1, 2, …, 100 or 101, 102, …, 200. But for the inter-BiLSTM the input time index is of non-continuous large-scale, e.g. time 1, 101, 201, …  or 2, 102, 202, … etc.. So the 3-D tensor needed to be permuted the second and third indexes for the input to the inter-BiLSTM. 
  That is to say, adjacent frames in the input of  the inter-BiLSTM are far apart and spread 
  across the global actual time dimension of the input mixed utterance.

  Although DPRNN-TasNet has achieved an amazing signal to distortion ratio improvement 
  (SDRi)~\cite{fevotte2005bss,vincent2006performance} in some public 
  data sets, there is a clear disadvantage in this structure, that is, all consecutive frames in the input of 
  inter-BiLSTM are far apart in the original utterance. 
  For the intra-BiLSTM, the input time index is the continuous natural one, e.g. time 1, 2, …, 100 or 101, 102, …, 200. But for the inter-BiLSTM the input time index is of non-continuous large-scale, e.g. time 1, 101, 201, … or 2, 102, 202, … etc.. That means for the input to inter-BiLSTM, the input time index is non-continuous, e.g. time 1, 101, 201, … etc.. These frames are far apart from in the utterance, and such kind of sequence may be used less frequently in speech processing.
  There are few sequence information and relationship 
  between the adjacent frames in the input of inter-BiLSTM. If the context information or mechanism 
  can be added to the 
  neighboring frames or to the structure of the inter-BiLSTM respectively, it is believed the performance will
  be improved.
  At the same time, in the training of DPRNN-TasNet, 
  the performance variance of different episodes is large, so some ensemble methods are tried to strengthen 
  DPRNN-TasNet. Also, the output of the DPRNN-TasNet can be refined again by combining the original mixed
  utterance to feed into the DPRNN-TasNet to result in better SDRi.
  These are the motivations for all the improvements in the 
  next section.

  \section{Speech separation with LaFurca}
  \label{sec:lafurca}

  The main work of this paper is to make several improvements to the dual-path BiLSTM module 
  (Figure~\ref{dprnn}) and 
  dual-path BiLSTM based framework (Figure~\ref{dprnn_pipeline}) for speech separation. 
  It should be noted that since DPRNN-TasNet is non-causal, so all our proposals are non-causal systems, please use with care.

  \begin{figure}
  \centering
  \hspace{-5mm}
  \includegraphics[width=1.0\linewidth]{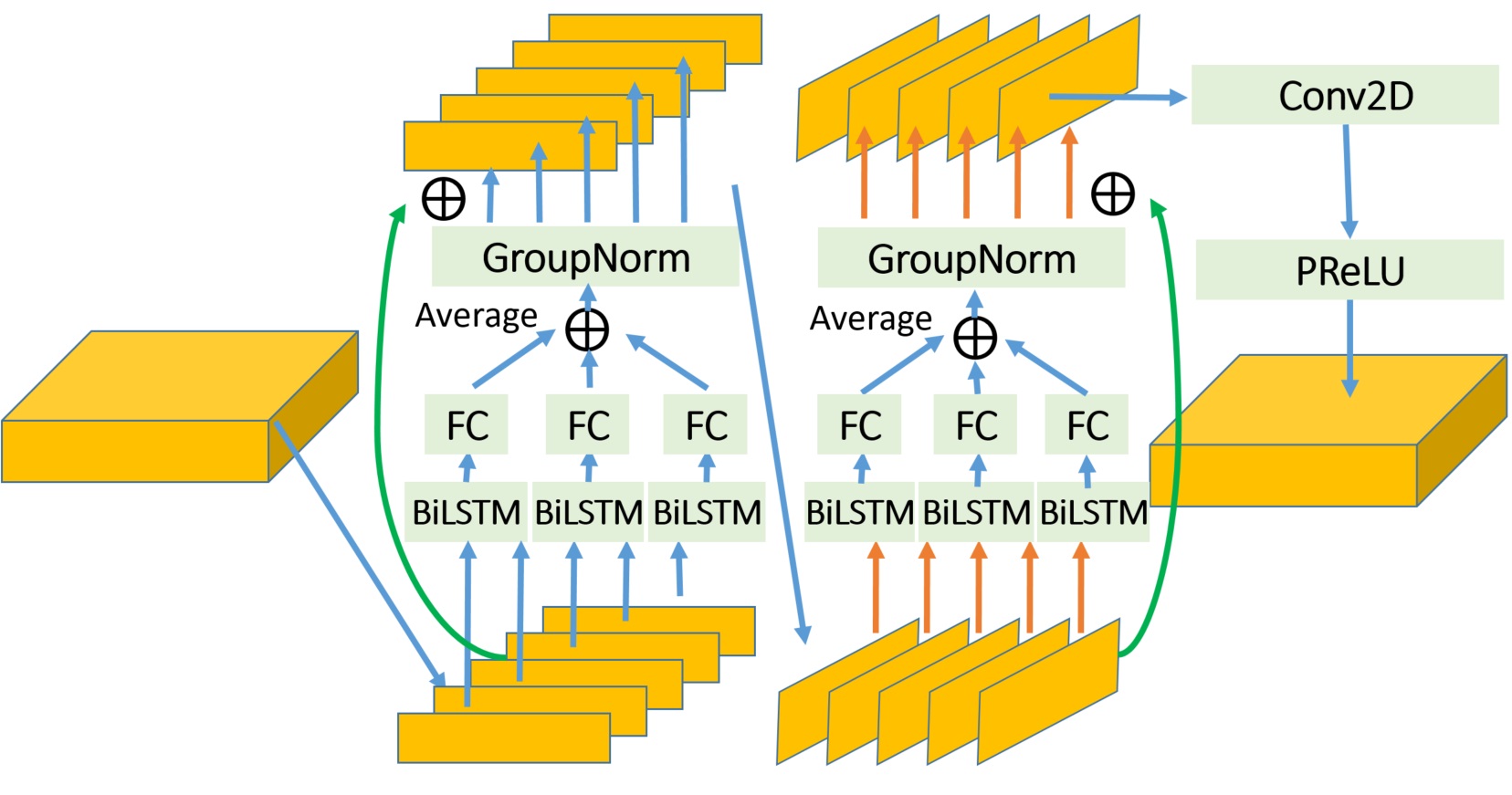}
  \hspace{-5mm}
  \caption{
  The structure of dual-path parallel BiLSTM.
  }
  \label{parallel_rnn}
  \end{figure}
  
  \subsection{Dual-path parallel BiLSTM}
   \label{sec:parallel_rnn}
  
  The performance of a single predictive model can always be improved by ensemble, that is to combine a set 
  of independently trained networks.
  The most commonly used method is to do the average of the model, which can at least  help to reduce the
   variance of the performance. As shown in Figure~\ref{parallel_rnn}, three identical parallel branches 
   are added in the intra-BiLSTM and inter-BiLSTM blocks respectively.
  The total output of each intra- and inter-parallel BiLSTM component is obtained by averaging the outputs 
  of all the different branches. 
  The same input is used for the parallel BiLSTMs. Although the BiLSTMs have same structure, their parameters are not the same, so the outputs are not the same.
  The reason why we do this 
  ensemble is to reduce the sub-variances of each block.
  Since there are 3 branches in this setting, the `sub-variances of each block' is to mean the performance variances between these 3 branches. If the 3 branches are tested individually, there will be differences between their performances. But if we put them together in parallel, then they will compensate each other and help each other to boost the performance.
  In our experiments, when a single branch is used, the performance of the network will fluctuate in different runs with different random seeds. That will result in variances in the separation effect. If we put several BiLSTM branches in parallel together, different branches can compensate each other in performing inference. That means if a certain branch results in poor prediction, other better branches can compensate for the poorer ones. Thus the performance of the ensemble parallel branches will be more stable even trained with different seeds.


  \subsection{Context-aware cross dual-path BiLSTM}
   \label{sec:cross_parallel_rnn}
  
\begin{figure}
    \centering
    \hspace{-5mm}
    \includegraphics[width=1.0\linewidth]{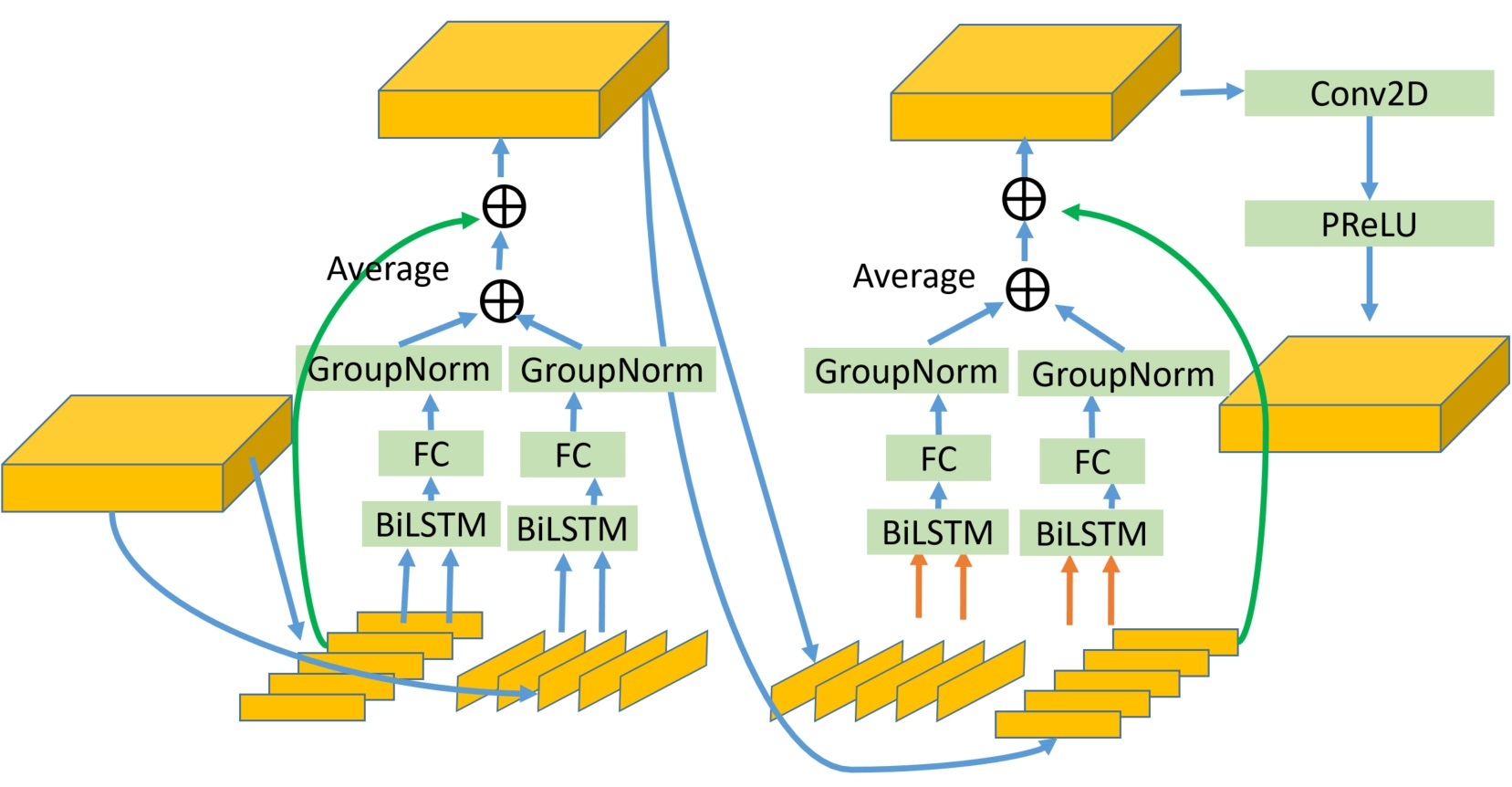}
    \hspace{-5mm}
    \caption{
    The structure of context-aware cross  dual-path BiLSTM.
    }
    \label{cross_parallel_rnn}
    \end{figure}

  Since the consecutive input frames of intra-BiLSTM are continuous in the original time axis, 
  intra-BiLSTM is more reasonable in modeling speech signals than inter-BiLSTM in DPRNN-TasNet. 
  Therefore, as shown in Figure~\ref{cross_parallel_rnn} we put intra-BiLSTM and inter-BiLSTM in parallel instead of 
  the original serial. 
  Their input differs in the arrangement of the data. 
  The outputs of intra-BiLSTM and inter-BiLSTM are averaged so that they can make use of
  global and local information from each other. In particular, inter-BiLSTM can make use of the 
  context information from the
  output of intra-BiLSTM to compensate for its weaknesses in this regard.

  \subsection{Iterative multi-stage refined dual-path BiLSTM for speech separation}
   \label{sec:iterative_la_furca}
   
\begin{figure*}[th]
  \centering
  \hspace{-5mm}
  \includegraphics[width=1.0\linewidth]{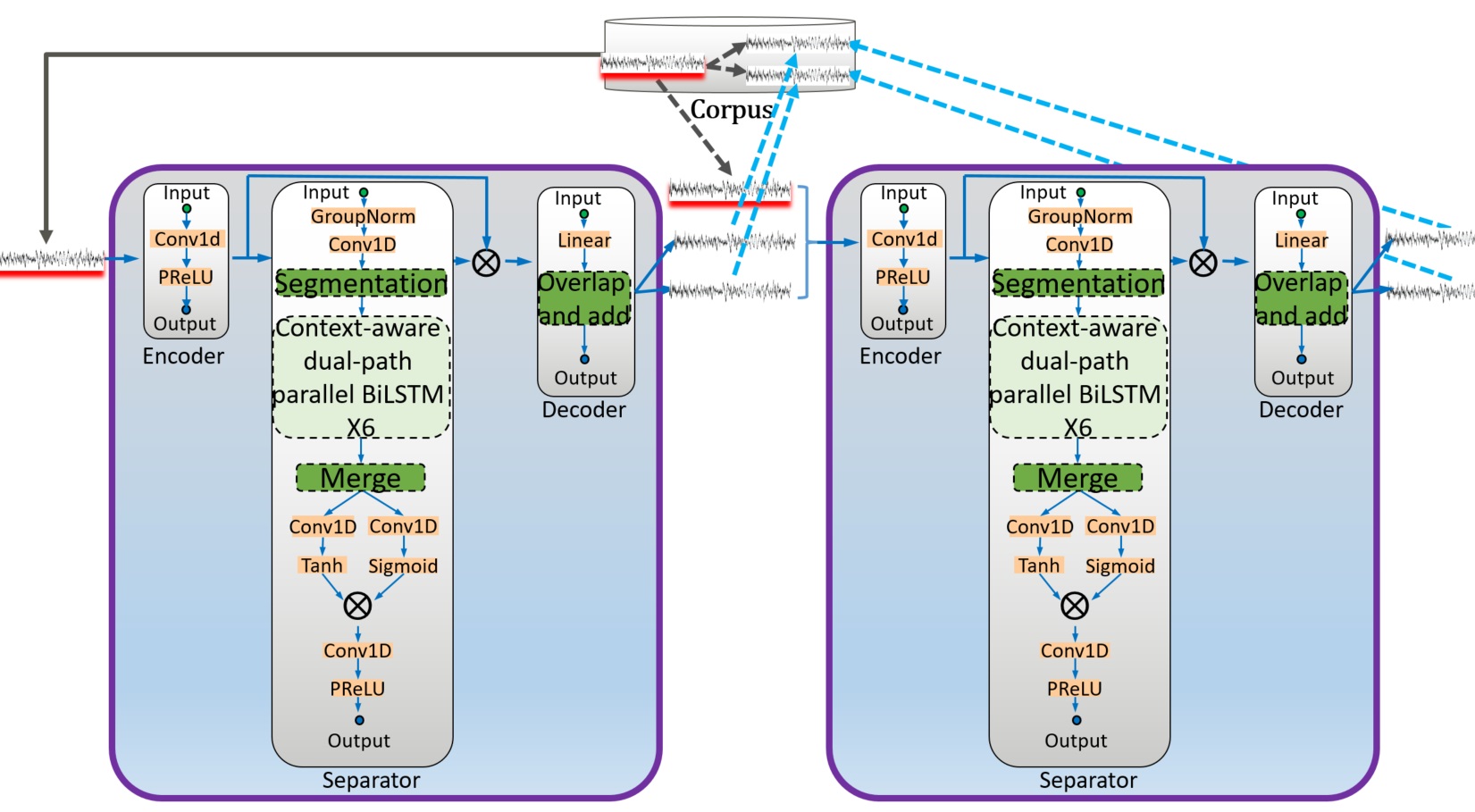}
  \hspace{-5mm}
  \caption{
  The structure of iterative  multi-stage refined context-aware dual-path parallel BiLSTM for speech separation, 
  which is also called LaFurca in this paper.
  }
  \label{iterative_la_furca}
  \end{figure*}

   Since the separated outputs and mixed input of the speech separation network must meet a consistent 
   condition, that is, 
   the sum of the separated outputs must be consistent with the mixed input. 
   Therefore, this consistent condition can also be used to refine the separated outputs of the network.
   Inspired by~\cite{isik2016single,kavalerov2019universal}, as shown in Figure~\ref{iterative_la_furca} we 
   propose to use a multi-stage iterative 
   network to do monaural speech separation. In each stage, there is a complete separate pipeline mentioned 
   earlier, such as any DPRNN-TasNet. The output of each stage pipeline is two separate utterances, 
   and these two utterances will be sent to the next stage sub-network along with the original mixed utterance
    to continue through the exact same pipeline, such as  DPRNN-TasNet, except that one of the input 
    dimension is tripled. 
   
  In our implementation and experiments, we tried different numbers of stages, including 2 stages and 3 stages. In other words, as shown in Figure~\ref{iterative_la_furca}, 2 or 3 TasNets with dual-path parallel BiLSTMs or context-aware cross dual-path BiLSTM, are connected in sequence to form an iterative refinement network. The insight we got was that 3 or more stages did not improve the performance anymore. That is, using only two stages is enough. When using three stages, the separation performances in SDR of the first stage and the second stage are the same, as can be seen from the Figure~\ref{iterative_loss}, in other words, one of the first two stages is not working. We will elaborate on this in the experimental part.
  
  \subsection{Combinations of different strategies}
  
\begin{figure}
    \centering
    \hspace{-5mm}
    \includegraphics[width=1.0\linewidth]{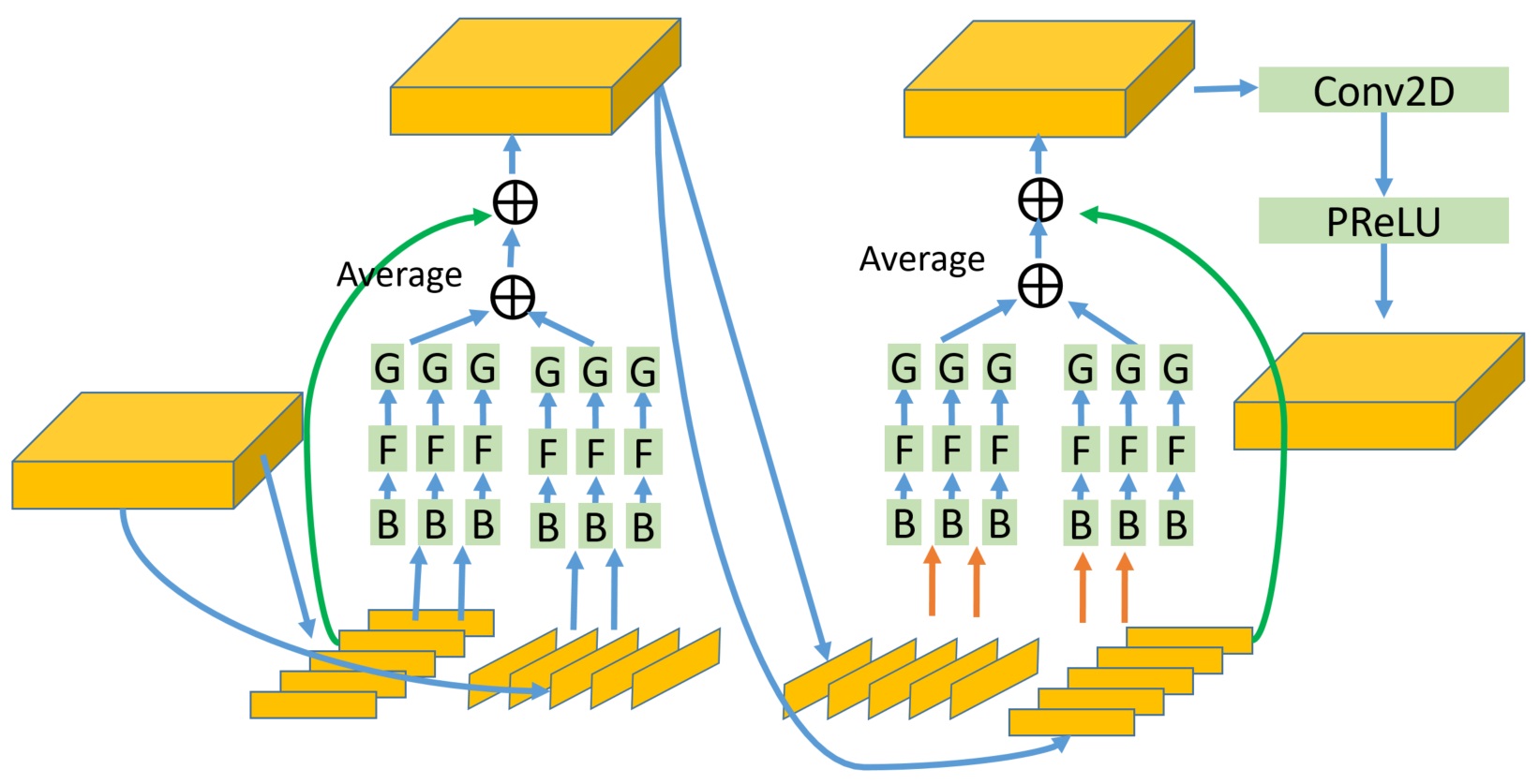}
    \hspace{-5mm}
    \caption{
    The structure of combined context-aware cross dual-path parallel BiLSTM. 	Here G, F, and B stand for the GroupNorm, FC, and BiLSTM respectively as the terms used in  Figure~\ref{cross_parallel_rnn}.
    }
    \label{combined_cross_parallel_rnn}
    \end{figure}

  Sections~\ref{sec:parallel_rnn}, ~\ref{sec:cross_parallel_rnn}, and~\ref{sec:iterative_la_furca} show three independent strategies to improve the original DPRNN-TasNet, and these three strategies can be combined in a free way, e.g.  we can put all three strategies in one big architecture. 
  Parallel  and context-aware BiLSTM in Figure~\ref{parallel_rnn}, ~\ref{cross_parallel_rnn} are are two extensions of the dual-path BiLSTM module in Figure~\ref{dprnn}, thus they can be the replacements of the dual-path BiLSTM module in the DPRNN-TasNet as shown in Figure~\ref{combined_cross_parallel_rnn}. That means parallel  and context-aware BiLSTM can also be used as modules of iterative multi-stage refined dual-path BiLSTM to enhance the single plain BiLSTM branch as shown in Figure~\ref{iterative_la_furca}. 
  Thus the only restriction on these combined strategies is the GPU memory. 
  
  \subsection{Utterance-Level Scale-Invariant SDR Objective Loss}
  \label{sec:loss}
  
  In this work, we directly use the scale-invariant signal-to-distortion ratio (SI-SDR)~\cite{roux2018sdr}, which is based on the most commonly used metrics SDR~\cite{fevotte2005bss,vincent2006performance} that is to evaluate the performance of source separation, as the training objective. SI-SDR measures the amount of distortion introduced by the output signal and define it as the ratio between the energy of the clean signal and the energy of the distortion.
  
  SI-SDR captures the overall separation quality of the algorithm. There is a subtle problem here. We first concatenate the outputs of  LaFurca  into a complete utterance and then compare with the input full utterance to calculate the SI-SDR in the utterance level instead of calculating the SI-SDR for one frame at a time. These two methods are very different in ways and performance. If we denote the output of the network by $s$, which should ideally be equal to the target source $x$, then SI-SDR can be given as~\cite{fevotte2005bss,vincent2006performance,roux2018sdr}
  \begin{equation*}
   \tilde{x}=\frac{\langle x , s \rangle}{\langle x , x \rangle} x, \quad e=\tilde{x}-s,\quad \text{SDR} = 10*\text{log}_{10}\frac{\langle \tilde{x} , \tilde{x} \rangle}{\langle e , e \rangle}.
  \end{equation*}
  Then our target is to maximize SI-SDR or minimize the negative SI-SDR as loss function respect to the $s$.
  
  \begin{figure}[th]
      \centering
      \hspace{-5mm}
      \includegraphics[width=1.0\linewidth]{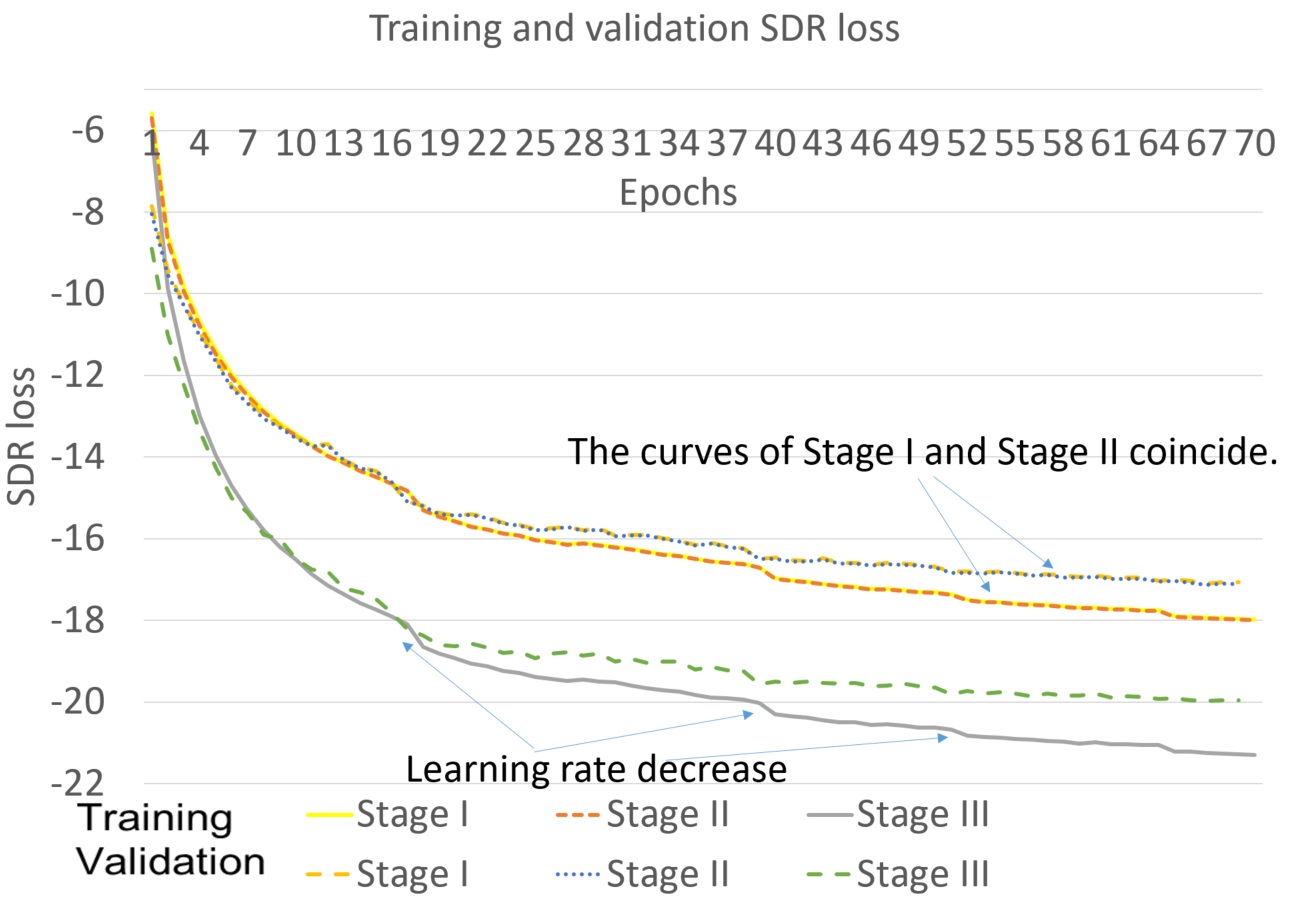}
      \hspace{-5mm}
      \caption{
      The losses of different stages from different epoch models on the training and validation data during the training of LaFurca.
      }
      \label{iterative_loss}
      \end{figure}
  
  To solve the tracing and permutation problem, the PIT training criteria~\cite{yu2017permutation} is employed in this work. We calculate the SI-SDRs for all the permutations, pick the maximum one, and take the negative as the loss. It is called the SI-SDR loss in this work.
  
  \begin{figure}[th]
  \centering
  \subfigure[Comparison of separation performance in SI-SDR (dB).]{
  \begin{minipage}[t]{0.5\linewidth}
  \centering
  \includegraphics[width=2.6in]{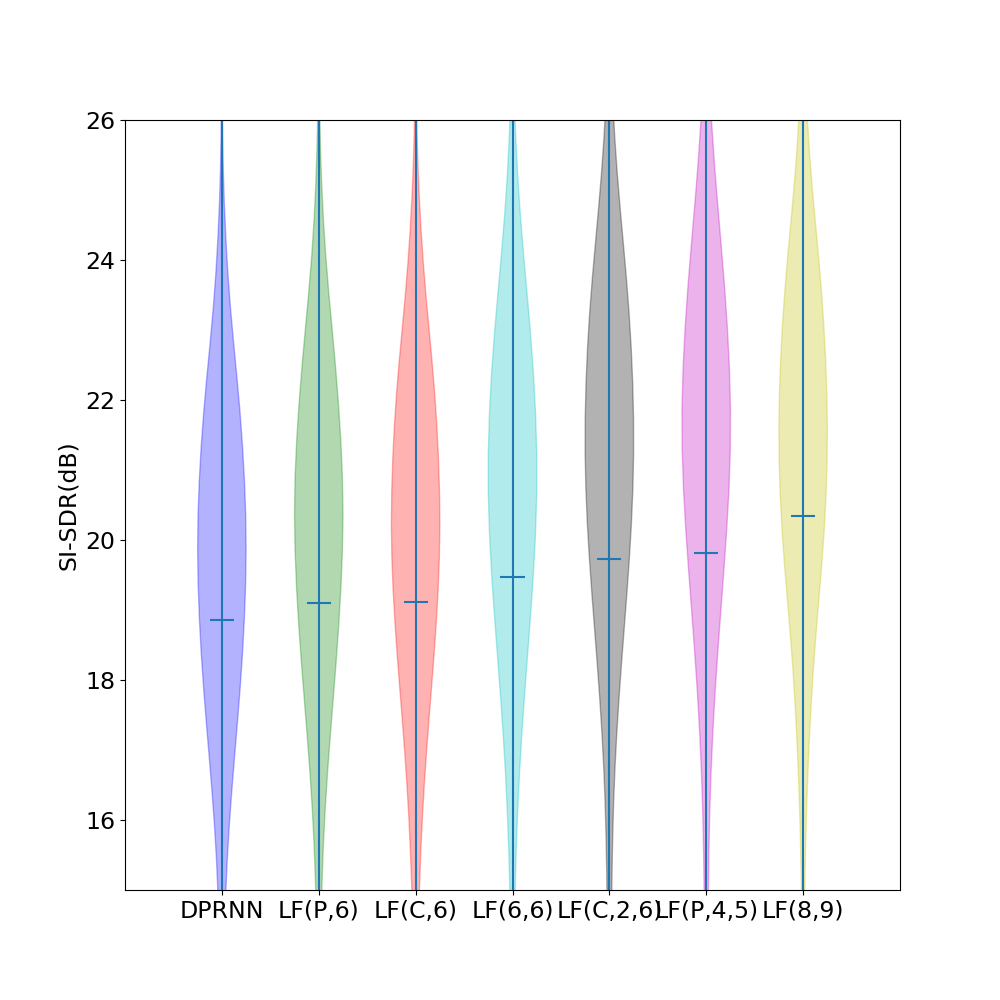}
  \label{sdr}
  \end{minipage}%
  }%
  \subfigure[Comparison of separation performance in SDR (dB).]{
  \begin{minipage}[t]{0.5\linewidth}
  \centering
  \includegraphics[width=2.6in]{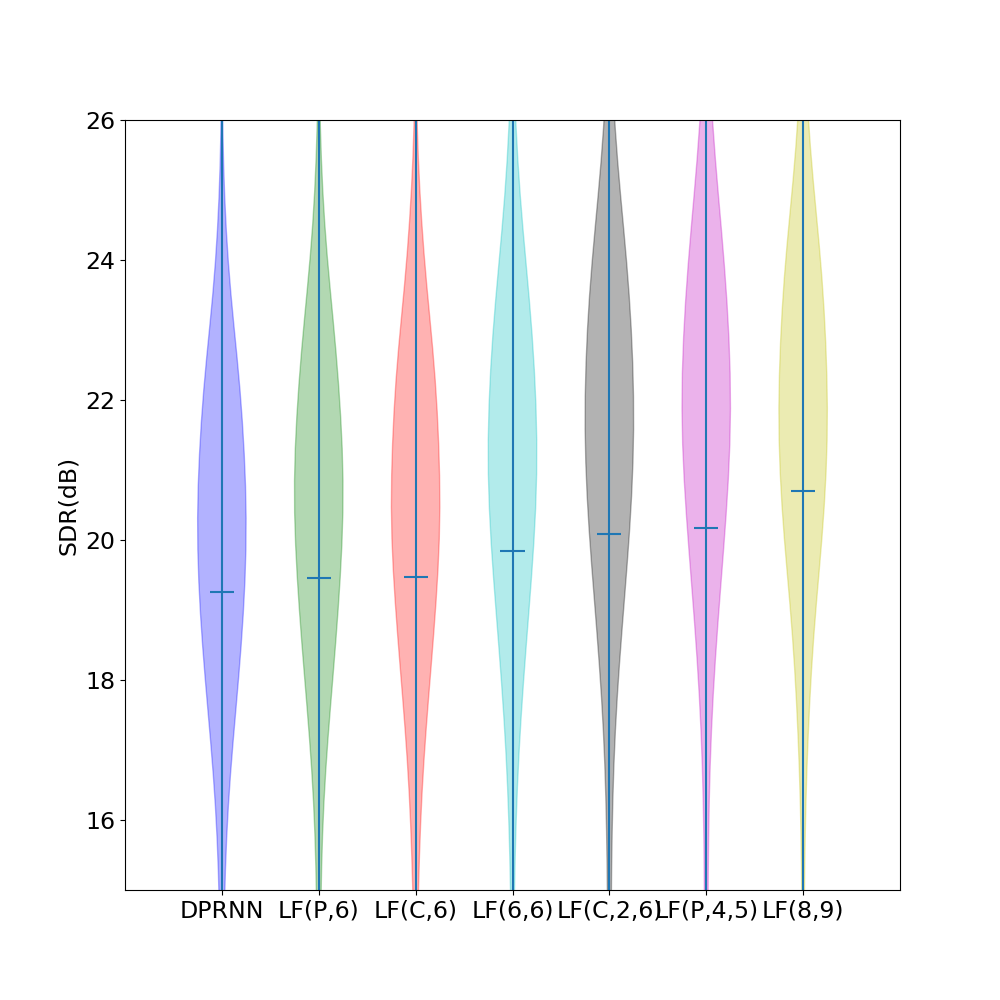}
  \label{pesq}
  \end{minipage}%
  }%
  
  \subfigure[Comparison of separation performance in PESQ.]{
  \begin{minipage}[t]{0.5\linewidth}
  \centering
  \includegraphics[width=2.6in]{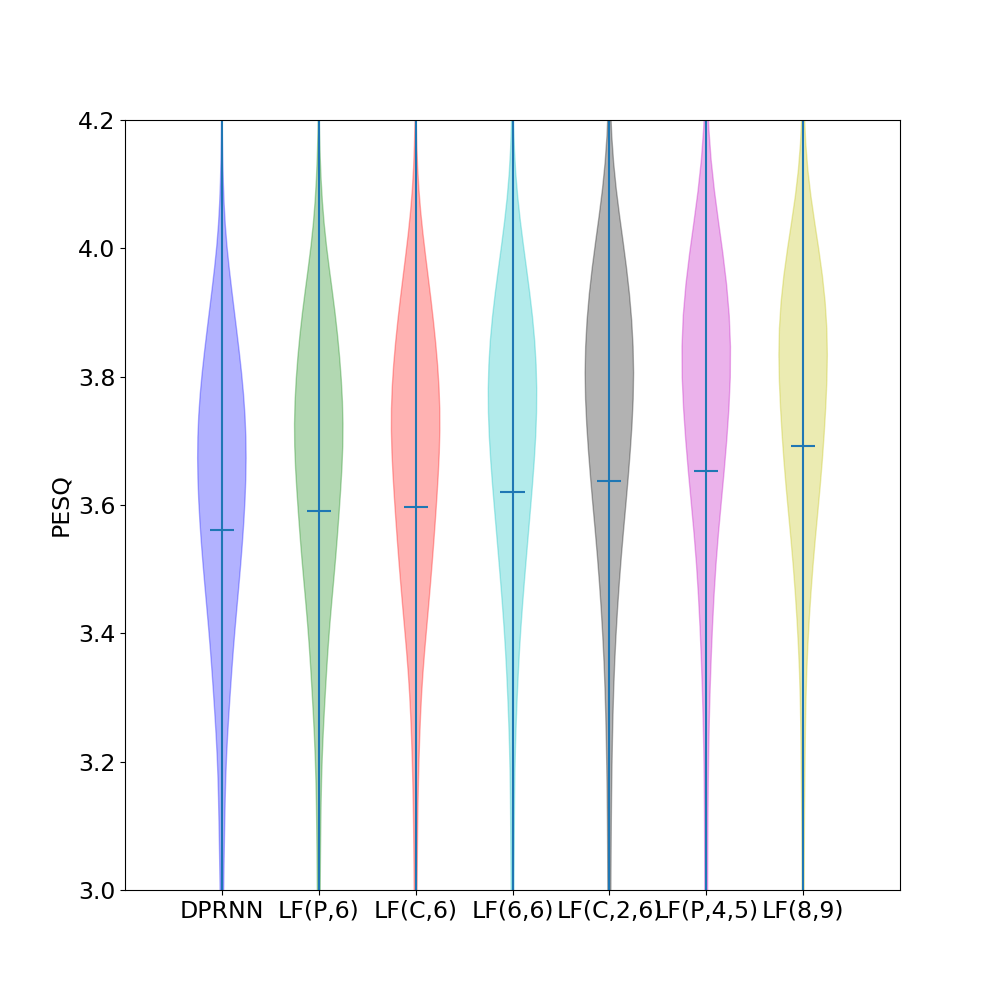}
  \label{pesq}
  \end{minipage}%
  }%
  \subfigure[Comparison of separation performance in ESTOI.]{
  \begin{minipage}[t]{0.5\linewidth}
  \centering
  \includegraphics[width=2.6in]{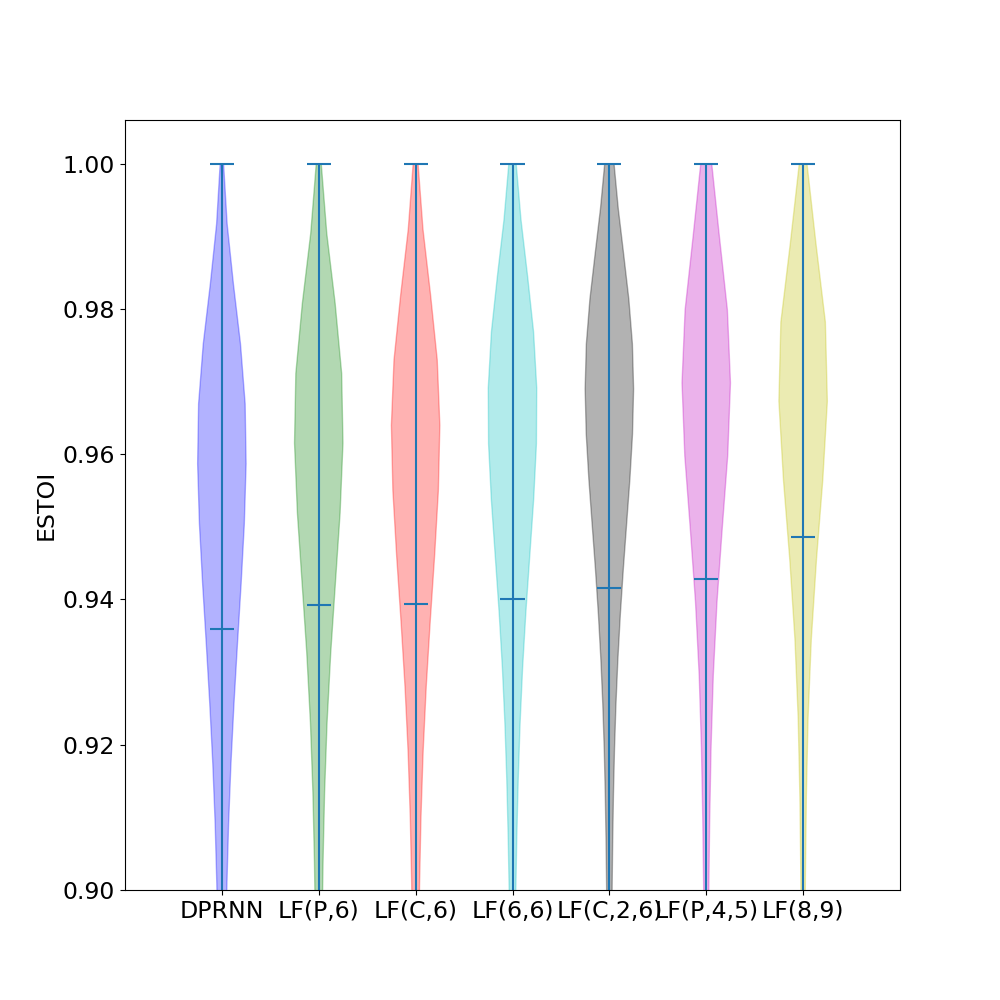}
  \label{estoi}
  \end{minipage}
  }%
  \centering
  \caption{Comparison of separation results in SI-SDR, SDR, PESQ, and ESTOI between DPRNN-TasNet and  LaFurcas on the WSJ0-2mix test set. \textbf{LF}  stands for  LaFurca.}
  \label{exp_results}
  \end{figure}
  
  The SI-SDR losses of the separated speech outputs at all stages with ground truth will be calculated, and then be averaged as the final loss.
  
  \subsection{Training}
  \label{sec:adam}
  
  During training Adam~\cite{kingma2014adam} serves as the optimizer to minimize the SI-SDR loss with an initial learning rate of 0.001 and scale down by 0.98 every two epochs. when the training loss increased on the development set, then restart training from the current best checkpoint with the halved initial learning rate. In other words, the learning rates of restart training are 0.001, 0.0005, 0.00025, etc. respectively. The batch size is set to 1 due to the limitation of the GPU memory size.
  
  \section{Experiments}
  \label{sec:experiments}
  
  \subsection{Dataset and neural network}
  \label{ssec:dataset}
  
  We evaluated our system on the two-speaker speech separation problem using the WSJ0-2mix dataset~\cite{hershey2016deep,isik2016single},  which is a benchmark dataset for two-speaker mono speech separation in recent years, thus most of those methods are compared on this dataset (In our earliest experiments, it has been shown that the results between two-talker and three-talker mixtures can be mutually confirmed. This is also been evidenced by many Luo et al.~\cite{luo2019conv-tasnet}  etc.. In other words, the case of two speakers can represent the case of the algorithm on 3 speakers.).
  WSJ0-2mix contains 30 hours of training and 10 hours of validation data. The mixtures are generated by randomly selecting 49 male and 51 female speakers and utterances in the Wall Street Journal (WSJ0) training set si\_tr\_s, and mixing them at various signal-to-noise ratios (SNR) uniformly between 0 dB and 5 dB (the SNRs for different pairs of mixed utterances are fixed by the scripts provided by~\cite{hershey2016deep,isik2016single} for fair comparisons). 5 hours of evaluation set is generated in the same way, using utterances from 16 unseen speakers from si\_dt\_05 and si\_et\_05 in the WSJ0 dataset. 
  
  In encoder and decoder, the window size is 2 samples and a 50\% stride size is used. The number of filters is set to be 64. As for the separator, 
  the number of dual-path BiLSTM in each stage is set to be 6, and with 128 hidden units in each direction of the BiLSTM.

  We evaluate the systems with the SI-SDRi~\cite{roux2018sdr}, the SDRi~\cite{fevotte2005bss,vincent2006performance}, perceptual evaluation of speech quality (PESQ)~\cite{rix2001perceptual} and extend short-time objective intelligibility (ESTOI)~\cite{jensen2016an} metrics used in~\cite{isik2016single,luo2018speaker, chen2017deep,liu2019divide,wang2019deep}. 
  The original SDR, that is the average SDR of mixed speech $y(t)$ with the original target speech $x_1(t)$ and $x_2(t)$ is 0.15. Table~\ref{tab:sdri} lists the results obtained by  LaFurca and almost all the results in the past four years, where IRM means the ideal ratio mask
  \begin{equation}
  M_s=\frac{|X_s(t,f)|}{\sum_{s=1}^{S}|X_s(t,f)|}
  \label{eq:irm}
  \end{equation}
  applied to the STFT $Y(t,f)$ of $y(t)$ to obtain the separated speech, which is evaluated to show the upper bounds of STFT based methods, where $X_s(t,f)$ is the STFT of $x_s(t)$.

  \subsection{Results and Discussions}
  \label{ssec:results}
  
  In this experiment, LaFurca is compared with several classical approaches, such as DPCL~\cite{hershey2016deep}, TasNet~\cite{luo2017tasnet}, Conv-TasNet~\cite{luo2018tasnet}, 
  and DPRNN-TasNet~\cite{luo2019dual}, Wavesplite~\cite{zeghidour2020wavesplit}, and Nachmani's~\cite{nachmani2020voice}. Use notation LaFurca(P, C, $x_1$, $x_2$, ... , $x_n$) to denote our prosposed system with dual-path \textbf{P}arallel BiLSTM, \textbf{C}ontext-aware dual-path  BiLSTM, and $x_1$ \textbf{c}ontext-aware dual-path \textbf{p}arallel BiLSTM blocks in the first stage, $x_2$ blocks in the second stage, etc.. If there is no 'P' or 'C'  in LaFurca, it uses ordinary dual-path BiLSTM. Thus DPRNN-TasNet is just LaFurca(6).

  \begin{table}[th]
  \caption[sdri]{SI-SDRi(dB), SDRi(dB), PESQ, and ESTOI(\%)  in a comparative study of different state-of-the-art separation methods on the WSJ0-2mix dataset.  \textbf{LF}  stands for  LaFurca.}\label{tab:sdri}
  \centering
  \begin{tabular}{c|c|c|c|c}
  \hline
  \hline
  Method & SI-SDRi & SDRi & PESQ & ESTOI\\
  \hline
  \hline
  DPCL~\cite{hershey2016deep}  & - & 5.9  & - & -\\
  uPIT-BLSTM~\cite{yu2017permutation}  & - & 10.0 & 2.84 & - \\
  ADANet~\cite{luo2018speaker}  & - & 10.5& 2.82 & - \\
  DPCL++~\cite{isik2016single} & - & 10.8 & - & - \\
  TasNet~\cite{luo2017tasnet} & - & 11.2 & - & -\\
  FurcaX~\cite{shi2019furcax} & - & 12.5 & - & - \\
  IRM & - &13.0 & 3.68  & 92.9 \\
  Wang et al.~\cite{wang2019deep}  & - & 15.4 & 3.45 & - \\
  Conv-TasNet~\cite{luo2019conv-tasnet}   & 15.3 & 15.6 & 3.24 & -\\
  Deep CASA~\cite{liu2019divide} & 17.7 & 18.0 & 3.51 & 93.2\\
  FurcaNeXt~\cite{zhang2020furcanext} & - & 18.4 & - & -\\
  DPRNN-TasNet~\cite{luo2019dual} & 18.8 & 19.0 & - & - \\
  Wavesplite~\cite{zeghidour2020wavesplit} & 19.0 & 19.2 & - & - \\
  Nachmani's~\cite{nachmani2020voice} & - & 20.12 & - & - \\
  \hline
  \hline
    \textbf{LF}(P, 6)  (ours)   & 19.09 &   19.31  & 3.59 & 93.92 \\
  \textbf{LF}(C, 6)  (ours)   & 19.11 &   19.33  & 3.60 & 93.93 \\
    \textbf{LF}(6 ,6)  (ours)   & 19.47 & 19.68 & 3.62 &  94.01 \\
   \textbf{LF}(C, 2, 6)  (ours)   & 19.73 &  19.93  & 3.64 & 94.16 \\
    \textbf{LF}(P, 4, 5)  (ours)   & 19.81 &  20.02  & 3.65 & 94.28 \\
    \textbf{LF}(8, 9)  (ours)   & 20.35 & 20.55 & 3.69 &  94.86 \\
  \hline
  \end{tabular}
  \end{table}

  Figure~\ref{iterative_loss} shows the losses of different stages from models of different epochs on the training and validation data during the training of a LaFurca(3,4,5). It can be seen that the SI-SDR obtained from the separated utterances of the first stage subnetwork and the separated utterances of the second stage subnetwork are almost coincident, whether it is on the training data or the validation data. That is to say, in practice, 2 stages are enough for LaFurca.
  
  Table~\ref{tab:sdri} lists the results obtained by our methods and almost all the results in the past four years, where IRM means the ideal ratio mask. Compared with these baselines, LaFurca obtained an absolute advantage, once again surpassing the performance of state-of-the-art. LaFurca has achieved the most significant performance improvement compared with baseline systems, and it breaks through the upper bound of STFT based methods a lot (more than 7.5dB).  
  
  For  the  \textbf{ablation} study,  Table~\ref{tab:sdri} shows that LaFurca(P, 6) is 0.3dB better than LaFurca(6) (which is the DPRNN-TasNet) in SDRi, LaFurca(C, 6) is 0.3dB better than LaFurca(6), and LaFurca(6, 6) is 0.6dB better than LaFurca(6) in SDRi.  That means the parallel BiLSTM, context-aware dual-path BiLSTM, and the iterative multi-stage refinement scheme are effective in boost the performance.
  Indeed in order to show the effectiveness of the proposed three novel strategies, LF(P,6), LF(C,6), and LF(6,6) are enough. This is indeed the ablation study. But if want to go further, a natural idea is that if the three strategies can be used together. It should have a stronger effect. So we tried LF(C,2,6) and LF(P,4,5), but due to the memory limitation of GPU, LF(C,6,6) and LF(P,6,6) cannot be trained on our GPU. LF(C,2,6), LF(P,4,5), and LF(8,9) are the largest scale we can achieve with limited GPU resources. At this time, the batch size has been set to 1. We see that with limited GPU memory resources, the limit we can achieve is LF(8,9), which also achieved the best performance in our experiments.
  
  Figure~\ref{exp_results} shows the comparison of separation results between DPRNN-TasNet and the  LaFurcas  on the WSJ0-2mix test set. It can be seen that most of our separated SI-SDRs and SDRs are concentrated above 16dB, and most of the utterances and overall are about 0.9dB higher than DPRNN-TasNet on average. Most of our separated PESQs are above 3.4 and most of our separated ESTOIs are above 0.94.

  \section{Conclusion}
  \label{sec:conclusion}
  
  In this paper, we investigated the effectiveness of dual-path BiLSTM block-based modeling for multi-talker monaural speech separation. We propose LaFurca to do speech separation. Benefits from the strength of end-to-end processing, parallel inter-intra data processing, context-aware dual-path BiLSTM, and the novel multi-stage refinement iterative scheme, the best performance of  LaFurca achieves the new state-of-the-art of 20.55dB SDRi on the public WSJ0-2mix data corpus.

  \section{Acknowledgment}
  We would like to thank Yi Luo at Columbia University  and Kaituo Xu at Beijing Kuaishou Technology for 
  sharing their implementations of Conv-TasNet and DPRNN block,
  and also valuable discussions on the training of DPRNN-TasNet.

\bibliographystyle{splncs03}
\bibliography{lafurca-arxiv}

\end{document}

%% file: macros.tex
\newcommand{\BALD}{\begin{aligned}}
\newcommand{\EALD}{\end{aligned}}
\newcommand{\BALDS}{\begin{aligned*}}
\newcommand{\EALDS}{\end{aligned*}}
\newcommand{\BCAS}{\begin{cases}}
\newcommand{\ECAS}{\end{cases}}
\newcommand{\BEAS}{\begin{eqnarray*}}
\newcommand{\EEAS}{\end{eqnarray*}}
\newcommand{\BEQ}{\begin{equation}}
\newcommand{\EEQ}{\end{equation}}
\newcommand{\BIT}{\begin{itemize}}
\newcommand{\EIT}{\end{itemize}}
\newcommand{\BMAT}{\begin{bmatrix}}
\newcommand{\EMAT}{\end{bmatrix}}
\newcommand{\BNUM}{\begin{enumerate}}
\newcommand{\ENUM}{\end{enumerate}}

\newcommand{\BA}{\begin{array}}
\newcommand{\EA}{\end{array}}

